\documentclass[conference]{IEEEtran}
\IEEEoverridecommandlockouts

\usepackage{cite}
\usepackage{amsmath,amssymb,amsfonts}
\usepackage{algorithmic}
\usepackage{graphicx}
\usepackage{multirow}
\usepackage{textcomp}
\usepackage{array}
\usepackage{tabularx}
\usepackage{xcolor}
\newcolumntype{C}{>{\centering\arraybackslash}X}

\def\BibTeX{{\rm B\kern-.05em{\sc i\kern-.025em b}\kern-.08em
    T\kern-.1667em\lower.7ex\hbox{E}\kern-.125emX}}
\begin{document}

\title{A Small-footprint Acoustic Echo Cancellation Solution for Mobile Full-Duplex Speech Interactions \\ 

}

\author{\IEEEauthorblockN{1\textsuperscript{st} Yiheng Jiang}
\IEEEauthorblockA{\textit{Speech Lab, Alibaba Group} \\
Beijing, China \\
jiangyiheng.jyh@alibaba-inc.com}
\and
\IEEEauthorblockN{2\textsuperscript{nd} Biao Tian}
\IEEEauthorblockA{\textit{Speech Lab, Alibaba Group} \\
Beijing, China \\
tianbiao.tb@alibaba-inc.com}
}

\maketitle

\begin{abstract}

In full-duplex speech interaction systems, effective Acoustic Echo Cancellation (AEC) is crucial for recovering echo-contaminated speech. This paper presents a neural network-based AEC solution to address challenges in mobile scenarios with varying hardware, nonlinear distortions and long latency. We first incorporate diverse data augmentation strategies to enhance the model's robustness across various environments. Moreover, progressive learning is employed to incrementally improve AEC effectiveness, resulting in a considerable improvement in speech quality. To further optimize AEC's downstream applications, we introduce a novel post-processing strategy employing tailored parameters designed specifically for tasks such as Voice Activity Detection (VAD) and Automatic Speech Recognition (ASR), thus enhancing their overall efficacy. Finally, our method employs a small-footprint model with streaming inference, enabling seamless deployment on mobile devices. Empirical results demonstrate effectiveness of the proposed method in Echo Return Loss Enhancement and Perceptual Evaluation of Speech Quality, alongside significant improvements in both VAD and ASR results.

\end{abstract}


\section{Introduction}
The performance of voice interaction systems is severely marred by acoustic echo~\cite{cite6,cite0}.
AEC is therefore a critical technology,
providing pristine audio communication by eliminating such undesirable feedback~\cite{cite1}.

\begin{figure*}[t]
    \centering
    \includegraphics[width=\textwidth]{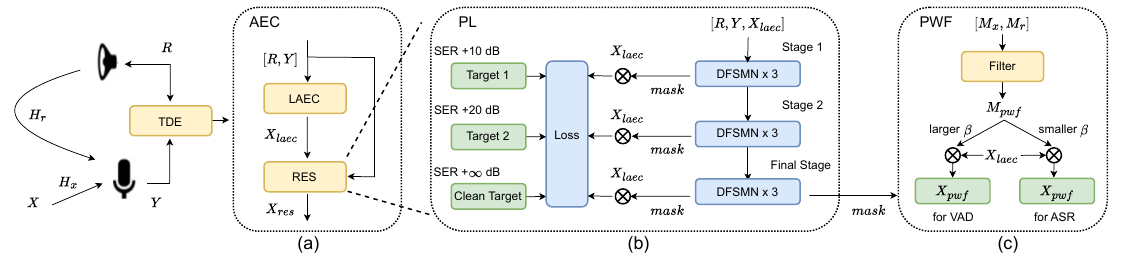} 
    \caption{Overview of (a) proposed AEC system comprising (b) RES model using Progressive Learning(PL) and (c) Post-processing with Wiener Filtering(PWF).}
    \label{fig_a2}
\end{figure*}

Recent studies on AEC, both with~\cite{cite_b14,cite_e11} and without~\cite{cite_e5,cite_e4} Neural Network (NN), have gained significant attention.
For NN-based AEC methods, a common approach involves two stages as depicted in Fig.~\ref{fig_a2}(a).
The first stage employs an adaptive filter to manage echo assumed to be linear, known as Linear AEC (LAEC)~\cite{cite3}.
\ \ \ \ \ \ \ \ \ \ \ \ \ \ \ \ \ \ \ \ \ \ \ \ \ \ \ \ \ \ \ \ \ \ \ \ \ \ \ \ \ \ \ \ \ \ \ \ \ \ \ \ \ \ \ \ \ \ \ \ \ \ \ \ \ \ \ \ \ \ \ \ \ \ \ \ \ \ \ \ \ \ \ \ \ \ \ \ \ \ \ \ \ \ \ \  The second stage incorporates NN-based techniques to further mitigate any residual and nonlinear echo, referred to as \ \ \ \ \ \ \ \ \ \ \ \ \ \ \ \ \ \ \ \ \ \ \ \ \ \ \ \ \ \ \ \ \  \ \ \ \ \ \ \ \ \ \ \ \ \ \ \ \ \ \ \ \ \ \ \ \ \ \ \ \ \ \ \ \ \  \ \ \ \ \ \ \ \ \ \ \ \ \ \ \ \ \ \ \ \ \ \ \ \ \ \ \ \ \ \ \ \ \ Residual Echo Suppressor (RES).
For instance, in~\cite{cite_b14}, an LAEC with multi-filter was used for echo cancellation, followed by an RES for subsequent echo suppression.
Additionally, within the two-stage framework, Wang et al.~\cite{cite_e11} applied multi-task learning to address echo suppression, noise reduction and near-end speech activity detection.

For AEC technology applied to full-duplex applications,
specifically when audio is output through a mobile phone's loudspeaker,
third-party developers face several challenges.
These include: ({\textit{a}}) device diversity and the resulting nonlinear distortions due to varying hardware characteristics~\cite{cite_a0},
\ \ \ \ \ \ \ \ \ \ \ \ \ \ \ \ \ \ \ \ \ \ \ \ \ \ \ \ \ \ \ \ \ ({\textit{b}}) the inconsistent effectiveness of built-in system-level AEC algorithms,
and ({\textit{c}}) variations latency between the reference and the microphone signal, ranging from a few to several hundred milliseconds~\cite{cite_a1}, occur due to hardware delays and software buffering~\cite{cite_e6}.

These challenges highlight the need for a flexible application-level AEC algorithm to supplement or cooperate with the built-in system-level AEC, thereby enhancing compatibility across various mobile devices and enabling effective full-duplex interactions.
In \cite{cite_a0}, the LAEC, combined with a statistical echo suppression method, was utilized in mobile phone Voice over IP (VoIP) scenarios. Nevertheless, it does not account for hardware differences among devices.
Additionally, Heitkaemper et al.~\cite{cite5} implemented a streaming AEC system to improve keyword spotting and ASR performance in smart voice assistants. However, this approach is limited to single interactions initiated by a wake word and does not address continuous full-duplex interactions, which require optimizing AEC with simultaneous consideration of both VAD and ASR effects.




In this paper, we propose a novel two-stage AEC system specifically designed for VAD and ASR tasks, intended for application in mobile full-duplex interaction scenarios.
Our contributions include:
({\textit{a}}) Utilizing multi-faceted Data Augmentation (DA) to enhance the model's adaptability across various mobile acoustic scenarios.
({\textit{b}}) Introducing a Progressive Learning (PL)~\cite{cite_d1,cite_d2} strategy into the RES training process, which is particularly effective in maintaining the fidelity of the speech signal.
({\textit{c}}) Applying a method using Post-processing with Wiener Filtering (PWF) to the RES outputs, with different tailored echo suppression parameters employed to optimize performance for VAD and ASR, respectively.\ \ \ \ \ \ \ \ \ \ \ \ \ \ \ \ \ \ \ \ \ \ \ \ 
({\textit{d}}) Prioritizing computational efficiency by designing a small-footprint model in a streaming manner, making it ideal for deployment on resource-constrained devices such as mobile phones.

\section{System}
\subsection{Problem Formulation}
For the AEC process in communication systems, the microphone signal \(y(n)\) is described as follows, assuming that the influence of background noise is ignored:
\begin{equation}
	y(n)=r(n) \ast h_r(n)+x(n) \ast h_x(n) \label{equation_a0}
\end{equation}
where \(n\) indexes a time sample, \(r(n)\) is loudspeaker signal (or far-end reference), \(x(n)\) represents target speech, \(h_r(n)\) and \(h_x(n)\) are convolutive acoustic transfer function~\cite{cite_d3}.
By transforming it into the time-frequency domain using Short-Time Fourier Transform (STFT), we can express it as follows:
\begin{equation}
	Y(t,f)=R(t,f)H_r(t,f)+X(t,f)H_x(t,f) \label{equation_a1}
\end{equation}
where \(t\) and \(f\) denote time frame and frequency bin index.
From here on, we will omit \((t,f)\) for simplifying the notation.

The AEC task aims to extract the reverberated target speech \(XH_x\) from the mixture signal \(Y\) when the reference \(R\) is available.
As illustrated in Fig.~\ref{fig_a2}(a), we first employ LAEC to eliminate linear echo component, with the output referred to as \(X_{laec}\).
Subsequently, with inputs comprising \(R\), \(Y\) and \(X_{laec}\),
the RES model is applied to further suppress remaining echo.
Additionally, TDE is an essential component introduced prior to the AEC system to address latency issues.


\subsection{Data Augmentation}

Prior research~\cite{cite_d0,cite_e6_2,cite_d4} has demonstrated that enhancing datasets through DA 
can lead to marked improvements in ASR and speech enhancement systems.
This suggests that DA should also confer benefits to AEC algorithm.
\subsubsection{\textbf{Reference augment}}
We apply SpecAugment~\cite{cite_d0}, which includes frequency masking and time masking, to the reference signal \(R\).
It is important to note that this operation is not applied to the mixture \(Y\).
Furthermore, following the idea in~\cite{cite5}, we randomly shift the reference ahead of the mixture signal by 0 to 20 ms after TDE alignment to simulate the latency between these two signals. This approach accounts for the fact that TDE does not always perfectly align the reference with the mixture signal during the inference phase.
The augmentation of the reference signal simulates temporal and spectral variabilities, thereby enhancing the network's robustness in recognizing the echo component, even when the correlation between the reference and the echo is relatively weak.


\subsubsection{\textbf{Merging Utterances}}

During the RES training phase, we synthesize mixture signals \(y(n)\) by randomly concatenating multiple utterances into a longer, continuous segment, which may include random overlaps or intervals between the selected utterances. These utterances are dynamically selected from dataset and may originate from different speakers. This approach effectively captures the complexities of overlapping and sequential speech patterns found in natural conversational environments.
Importantly, the echo within this longer mixture segment is derived from the same recording, without any merging operations, thereby simulating more authentic successive interactions.
\subsection{Progressive Learning}

Traditional deep learning approaches for speech enhancement typically process noisy spectral inputs to produce clear outputs. However, accurately mapping these complex relationships within NN is challenging, and the functioning of the network's intermediate layers remains unclear and elusive.

The strategy of PL offers a novel solution to this elusive problem~\cite{cite_d1}
by segmenting the layers of NN into several stages.
Each stage builds upon the output of its predecessor,
targeting the spectral of speech with a progressively higher Signal-to-Noise Ratio (SNR).
This staged approach not only makes the incremental SNR enhancements across layers transparent
but also emphasizes the recovery of clear speech signals at every stage.


The concept of PL can also be extended to the AEC task. As illustrated in Fig.~\ref{fig_a2}(b), we guide the RES model to progressively eliminate echo by increasing the Signal-to-Echo Ratio (SER) of the target signal at each stage. 
Note that the final target is clean speech without any interference from echo.


\subsection{Post-Processing}

In full-duplex systems, both VAD and ASR depend on effective AEC algorithms. Our proposed PWF post-processing technique, applied after the RES, generates two distinct signals, each specifically tailored and sent to VAD and ASR to meet their requirements, as illustrated in Fig.~\ref{fig_a2}(c).

For the RES training phase, we input \([R,Y]\) into LAEC to produce output \(X_{laec}\).
Simultaneously, \([RH_r,R]\) is feed into LAEC to obtain the residual echo \(R_{laec}\). Assuming that LAEC does not introduce any speech distortion, it follows that theoretically:
\begin{equation}
	X_{laec} = XH_x + R_{laec}\label{equation_a2}
\end{equation}
The output \(X_{laec}\) serves as an input to the RES model, while \(XH_x\) and \(R_{laec}\) are training targets.
Our RES model generates real-valued time-frequency masks \(M_x\) and \(M_r\) to predict the target speech and residual echo, respectively, using sigmoid function to ensure their values are in the range of 0 to 1.

Assuming that speech and echo are statistically independent, we can express the solution of the Wiener filtering in the time-frequency domain as follows~\cite{cite_e7}:
\begin{equation}
    \begin{split}
	M_{pwf} &= \frac{P_{zx}}{P_{zz}} = \frac{P_{xx}}{P_{zz}} \\[0.5em]
    &= \frac{{\mid M_xX_{laec}\mid}^2}{{\mid M_xX_{laec} + M_rX_{laec}\mid}^2} \\[0.5em]
    &= \left({\frac{M_x}{M_x+M_r}}\right)^2 \label{equation_a3}
    \end{split}
\end{equation}
where \(P_{xx}\) and \(P_{zz}\) are the power spectral densities of the predicted speech and mixture, respectively, and \(P_{zx}\) is the cross spectral density between these two signals, all of which are computed at the current time frame. \(M_xX_{laec}\) and \(M_rX_{laec}\) represent the predictions of target speech \(XH_x\) and target echo \(R_{laec}\), respectively. The final output of the post-processing step is defined as:
\begin{equation}
	X_{pwf} = M_{pwf}^{\ \ \ \ \beta} X_{laec} \label{equation_a4}
\end{equation}

where \(\beta\) is an additional exponent. During the inference phase, \([R,Y]\) is sent to the AEC system to generate \(M_x\) and \(M_r\), After this, the PWF process is applied to compute the final output \(X_{pwf}\).
By adjusting the parameter \(\beta\), we can modulate the echo cancellation effect.
A smaller \(\beta\) for ASR may result in more residual echo being retained, but facilitates better preservation of speech quality.
Conversely, a larger \(\beta\) enhances echo suppression for VAD, enabling more accurate identification of the start and end points in speech recordings, even though it may introduce some distortion.
This approach enables the simultaneous optimization of both ASR and VAD with a negligible increase in computational complexity.

\subsection{Model Structure and Loss Function}
In our study, the LAEC was implemented following~\cite{cite_e11}, and we employed the Deep Feedforward Sequential Memory Network (DFSMN)~\cite{cite_b13} as the backbone architecture of the RES model, with Fully Connected layer (FC) + sigmoid activation for mask prediction.
Furthermore, we included \ \ \ \ \ \ \ \ \ \ \ \ \ \ \ \ \ \ \ \ \ \ \ \  \ \ \ \ \ \ \ \ \ \ \ \ \ \ \ \ \ \ \ \ \ \ \ \  \ \ \ \ \ \ \ \ \ \ \ \ \ \ \ \ \ \ \ \ \ \ \ \ FC + sigmoid layers at each intermediate stage to produce additional masks for PL training.
To ensure streaming inference capabilities along with effective context memory, each layer only allowed for a 20-frame lookback, explicitly excluding any future frames.
As depicted in Fig.~\ref{fig_a2}(b), we divided the model into three stages, each containing three DFSMN layers.
The hidden size of DFSMN was 128, resulting in a total of 432k parameters in the RES model, which ensured lightweight and computationally efficient design suitable for mobile deployment.

The loss function comprised a weighted sum of Modulation Loss~\cite{cite_d6}, SNR Loss~\cite{cite_d8} 
and PMSQE Loss~\cite{cite_d7}, with respective weights of 0.1, 0.9, and 10.
These weights were chosen to ensure that the numerical values of the different loss components are approximately equal,
maintaining a balance across them.



\section{Experiments}

\subsection{Data Preperation}

Our study focuses on optimizing VAD and ASR performance in mobile full-duplex interactions, conducting experiments using mobile phones and potentially applicable to other mobile devices. However, there are currently no publicly available AEC datasets designed for mobile scenarios with the necessary annotations for simultaneous VAD and ASR testing. This leads us to record and construct an internal dataset for our experiments.
We collected echo recordings and their corresponding reference signals from 100 commonly used smartphones, with each device providing around 30 minutes of continuous recordings.
Clean speech and noise clips were sourced from the DNS challenge~\cite{cite_e0}.
The Room Impulse Response (RIR) was generated using gpuRIR~\cite{cite_e9}, with randomly selected reverberation times (RT60) between 0.1 and 0.8 s.
The audio data was sampled at 16 kHz, and the RES model utilized STFT as its input feature, with a frame length of 40 ms and a frame shift of 20 ms.

For the Echo Return Loss Enhancement (ERLE) metric, a portion of recorded echoes served as a far-end single-talk dataset prior to training.
For the Perceptual Evaluation of Speech Quality (PESQ)~\cite{cite_e8} metric, these test echoes were mixed with DNS challenge speech clips for synthesized double-talk evaluation. 
Additionally, VAD and ASR were tested using a dataset of real recorded audio. 
This recorded dataset comprises 40 mobile phones, each capturing utterances at two loudspeaker volume levels (70\% and 100\%) in noisy double-talk environments, totaling approximately 4,000 utterances. There are around 400 segments, each containing 10 utterances, with a few seconds of interval between each adjacent utterance to facilitate VAD testing.

\subsection{Evaluation Results}
\subsubsection{\textbf{PESQ and ERLE}}
Table \ref{tab_b0} presents the PESQ outcomes for the synthesized double-talk dataset and the ERLE outcomes for the far-end single-talk dataset. The SER levels for the double-talk data are configured at [-20, -10, 0, 10] dB. This configuration is based on our observation that commonly used mobile phone recordings typically exhibit around -20 dB \ \ \ \ \ \ \ \ \ \ \ \ \ \ \ \ \ \ \ \ \ \ \ \ \ \ \ \ \ \ \ \ \ \ \ \ \ \ \ \ \ \ \ \ \ \ \ \ \ \ \ \ \ \ \ \ \ \ \ \ \ \ \ \ \ \ \ \ \ \ \ \ \ \ \ \ \ \ \ \ \ \ \ \ \ \ \ \ \ \ \ \ \ \ \ \ \ \ \ \ \ \ \ \ when the volume is set to 100\%.


The two-stage AEC in Table \ref{tab_b0} serves as our baseline and does not incorporate any optimizations introduced. 
The PESQ and ERLE results indicate that incorporating DA technique significantly enhances AEC performance by improving the adaptability of the RES model to diverse mobile acoustic environments.
Additionally, the ERLE results demonstrate that PL may not significantly enhance echo suppression. 
However, PL is an effective strategy for enhancing speech quality particularly in low SER scenarios. At -20 dB, the AEC employing DA + PL approach achieves the highest PESQ score of 1.83, compared to 1.57 for the system without PL method.

\begin{table}[]
    \centering
    \caption{Comparison of PESQ on synthesized double-talk dataset and ERLE(DB) on far-end single-talk dataset.} \label{tab_b0}
    \begin{tabularx}{0.45\textwidth}{l|ccCc|c}
    \hline
    \multirow{2}{*}{Method} & \multicolumn{4}{c|}{PESQ$ \ \uparrow $}                                                                                   & \multicolumn{1}{c}{ERLE$ \ \uparrow $} \\
                            & \multicolumn{1}{c}{-20 dB} & \multicolumn{1}{c}{-10 dB} & \multicolumn{1}{c}{0 dB} & \multicolumn{1}{c|}{10 dB} & \multicolumn{1}{c}{-}    \\ \hline
                            two-stage AEC                     & 1.26                      & 1.95                      & 2.51                    & 2.77                      & 35.12                    \\
    \ +   DA                  & 1.57                      & 2.09                      & 2.66                    & \textbf{2.80}                     & 41.46                    \\
    \ +   DA + PL             & \textbf{1.83}                      & \textbf{2.27}                      & \textbf{2.67}                    & 2.78                       & \textbf{42.77}                   \\ \hline
    \end{tabularx}
    \end{table}

\begin{table}[]
    \centering
    \caption{Comparison of VAD and ASR on a real recorded mobile phone dataset, evaluating VAD by DCF(\%), and ASR by WER(\%).} \label{tab_b1}
    \begin{tabularx}{0.45\textwidth}{l|CC|CC}
    \hline
    \multirow{2}{*}{Method} & \multicolumn{2}{c|}{VAD-DCF$ \ \downarrow $} & \multicolumn{2}{c}{ASR-WER$ \ \downarrow $} \\
                            & Vol.70        & Vol.100       & Vol.70       & Vol.100       \\ \hline
                            two-stage AEC                     & 5.30         & 8.78          & 10.76        & 20.24         \\
    \ + DA                  & 2.41         & 5.55          & 9.24         & 17.91         \\
    \ + DA + PL             & 2.17         & 4.81          & 8.83         & 15.81          \\
    \ + DA + PL + two masks            & 2.35         & 4.64          & 8.70         & 15.49         \\
    \ + DA + PL + PWF       & \textbf{1.73}         & \textbf{3.68}          & \textbf{7.05}         & \textbf{11.72} \\ \hline
    \end{tabularx}
    \end{table}

\subsubsection{\textbf{VAD and ASR}}
In this evaluation,
the VAD algorithm utilizes semantic-VAD~\cite{cite_e2},
while the ASR system employs Paraformer~\cite{cite_e3}.
Table \ref{tab_b1} presents a comparison of VAD metric (Detection Cost Function, DCF) and ASR metric (Word Error Rate, WER) based on a real recorded mobile phone dataset.
DCF is defined based on the measures from~\cite{cite_e2}, focusing on two key indicators: false triggers \(P_{false}\) and missed detections \(P_{miss}\). It is calculated as \ \ \ \ \ \ \ \ \ \ \ \ \ \ \ \ \ \ \ \ \ \ \ \ \ \ \ \ \ \ \ \ \ \ \ \ \ \ \ \ \ \ \ \ \ \ \ \ \ \ \ \ \ \ \ \ \ \ \ \ \(DCF=0.75P_{false}+0.25P_{miss}\), placing greater emphasis on \(P_{false}\) due to its greater impact on user experience in full-duplex interactions.
Vol.70 and Vol.100 refer to the loudspeaker volume levels set at 70\% and 100\%, respectively. The notation ``two masks'' indicates that the RES model produces two masks, namely \(M_x\) and \(M_r\). However, in this context, only \(M_x\) is utilized without applying Wiener filtering. In contrast, the PWF approach, as outlined in this table, processes both masks through Wiener filtering and uses \(M_{pwf}\) for the final output.

As shown in Table \ref{tab_b1}, the combination of DA and PL techniques leads to a significant improvement, and further integration of PWF yields the best performance.
Specifically, the method that includes PWF achieves the lowest DCF values of 1.73 for Vol.70 and 3.68 for Vol.100, respectively, as well as the lowest WER values of 7.05 for Vol.70 and 11.72 for Vol.100. This underscores the efficacy of DA, PL, and PWF techniques in improving VAD and ASR capabilities under mobile scenarios.
The results using ``two masks" are comparable to those of DA + PL method (without predictions of two masks), indicating that training exclusively with two masks, in the absence of PWF process, offers no significant improvements.

\subsubsection{\textbf{Different outputs in PL}}
We conducted a systematic ASR evaluation of the PL framework based on the DA + PL experiment. The results in Table~\ref{tab_b2} show that PL incrementally enhances ASR accuracy through the intermediate stages to the final stage. As in Fig.~\ref{fig_a2}(b), the RES model was designed to increase the SER by 10 dB at each middle stage. This approach resulted in training targets of [+10, +20, +\(\infty\)] dB, with +\(\infty\) representing the echo-free, clean target for the network's final output.

It is noteworthy that, in our experiments, employing intermediate outputs for ASR system, as suggested in previous work~\cite{cite_d1}, dose not yield optimal results.
This may be attributed to the challenging mobile scenarios (with a SER around -20 dB at 100\% volume), where our small-footprint RES model struggles to effectively address these conditions using only intermediate layers.

\subsubsection{\textbf{Different parameters in PWF}}
Table \ref{tab_b4} highlights the importance of utilizing PWF with different \(\beta\) parameters to meet the specific needs of both VAD and ASR tasks, with optimal values of 0.6 for VAD and 0.2 for ASR.
However, the optimal \(\beta\) value for PESQ is 0.4, indicating that improved speech enhancement scores do not always result in lower WER~\cite{cite_e10}.
Additionally, as \(\beta\) increases, ERLE consistently rises. Nevertheless, this increase in echo reduction does not guarantee improved VAD and ASR performance, as it may also lead to more speech distortion.

\begin{table}[]
    \centering
    \caption{Comparison different outputs in PL in terms of WER (\%).} \label{tab_b2}
    \tabcolsep=0.6cm 
    \begin{tabularx}{0.45\textwidth}{l|CC}
        \hline
        Layer        & Vol.70  & Vol.100 \\
        \hline
        Stage 1 (+10 dB)        & 17.65 & 53.24 \\
        Stage 2 (+20 dB)        & 10.32 & 22.64 \\
        Final Stage (+\(\infty\) dB) & \textbf{8.83}  & \textbf{15.81} \\
        \hline
        \end{tabularx}
    \end{table}

\begin{table}[]
    \centering
    \caption{Comparison of different \(\beta\) value in PWF} \label{tab_b4}
    \begin{tabularx}{0.45\textwidth}{l|ccCC}
    \hline
    \(\beta\)  & VAD-DCF $ \ \downarrow $          & ASR-WER $ \ \downarrow $          & PESQ $ \ \uparrow $         & ERLE $ \ \uparrow $          \\ \hline
    0.1 & 29.72         & 12.24         & 2.18          & 14.09          \\
    0.2 & 10.56         & \textbf{9.39} & 2.33          & 25.58          \\
    0.4 & 4.45          & 10.04          & \textbf{2.39}          & 38.24          \\
    0.6 & \textbf{2.70}          & 13.13         & 2.16          & 43.11          \\
    0.8 & 3.19          & 17.95         & 1.81          & \textbf{45.93} \\ \hline
    \end{tabularx}
    \end{table}

\section{Conclusion}

Our study introduces a novel AEC approach to address the challenges in mobile full-duplex interactions. By developing a small-footprint streaming RES model that leverages DA, PL, and PWF techniques, we achieve significant improvements in PESQ and ERLE, as well as enhanced performance in downstream VAD and ASR tasks. The integration of DA enhances the adaptability of the RES model to diverse acoustic environments, while PL ensures effective enhancement of speech quality through a progressive learning framework. Additionally, PWF enables customized echo suppression parameters to meet the differing needs of VAD and ASR.

\end{document}